\documentclass[twocolumn]{article}
\usepackage[margin=2cm]{geometry}

\usepackage{authblk}

\title{Matrix Representation of Arbitrarily Controlled Quantum Gates}
\author[1]{Marco Lewis\footnote{m.j.lewis2@newcastle.ac.uk}}
\author[1]{Sadegh Soudjani}
\author[1]{Paolo Zuliani}
\affil[1]{Newcastle University}

\usepackage[utf8]{inputenc}
\usepackage[english]{babel}
\usepackage{amsfonts}
\usepackage{amsthm}
\usepackage{physics}
\usepackage{caption}
\usepackage{subcaption}
\usepackage{tikz}
\usepackage{hyperref}
\usetikzlibrary{quantikz}
\usepackage{comment}
\usepackage[nocompress]{cite}

\newcommand{\freezecache}{frozencache,cachedir=styminted}

\usepackage[\freezecache]{minted}
\RecustomVerbatimEnvironment{Verbatim}{BVerbatim}{}
\usepackage{setspace}

\usepackage{lineno}
\newtheorem{theorem}{Theorem}[section]
\newtheorem{corollary}{Corollary}[theorem]
\newtheorem{lemma}[theorem]{Lemma}

\theoremstyle{definition}
\newtheorem{definition}{Definition}[section]
\newtheorem{exmp}{Example}[section]

\theoremstyle{remark}

\newcommand{\twodots}{\mathinner {\ldotp \ldotp}}
\newcommand{\bs}[1]{\{0,1\}^{#1}}
\newcommand{\bsn}{[0 \twodots N)}
\newcommand{\hil}{\mathcal{H}}
\newcommand{\hilsys}{\hil_\text{sys}}
\newcommand{\fdomain}{f^{-1}(1)}

\def\ie{{\em i.e., }}

\newcommand{\appendproofs}{1}

\begin{document}
\maketitle

\begin{abstract}
    Controlled operations allow for the entanglement of quantum registers. In particular, a controlled-$U$ gate allows an operation, $U$, to be applied to the target register and entangle the results to certain values in the control register. This can be generalised by making use of the classical notion of conditional statements, where if a value (or state) satisfies some condition then a sequence of operations can be performed.
    A method is introduced to represent these generalised controlled operations that are based on classical conditional statements. Throughout examples are given to highlight the use of introduced gates.
\end{abstract}

\section{Introduction\label{sec:intro}}
In classical computing, conditional (if-then-else) statements are a commonly used paradigm within programming languages to affect the control and execution of a program.

In quantum computing, the conditional flow of a program is normally captured through the use of controlled operations. The simplest example is the CNOT gate that, when given a control and a target quantum bit (qubit), will flip the state of the target qubit if the control qubit is in the $\ket{1}$ state, as shown in Figure~\ref{fig:cnot}. Quantum unitary operations (represented by $U$) can be controlled by decomposing the controlled-$U$ gate into CNOTs and single unitary gates \cite{ElementaryGates}. However, it is not possible to decompose a controlled-$U$ gate if $U$ is unknown \cite{Thompson13, Araujo13, Soeda13}; the conditions that make quantum operations controllable were shown in \cite{Bisio15}.

\begin{figure}[ht]
     \centering
     \begin{quantikz}
     \lstick{$\alpha \ket{0} + \beta \ket{1}$} & \ctrl{1} & \qw \rstick[wires=2]{$\alpha \ket{0}\ket{0} + \beta \ket{1}\ket{1}$}\\ 
     \lstick{$\ket{0}$} & \targ{} & \qw
     \end{quantikz}
     \caption{A diagram of the CNOT gate. The black dot on the top wire represents a control on the $\ket{1}$ state.}
     \label{fig:cnot}
\end{figure}

Various notions of conditional statements have been suggested in different programming languages.
The \textbf{qwhile} language is a quantum extension of the while-language, with two different forms of conditional statements. In \cite{Ying11}, a classical conditional statement is used in the \textbf{qwhile} language, where only classical variables can affect the control of a program (although one could implement conditional operators as a unitary gate). However, in works such as \cite{YingFHL, Liu17}, conditional statements are based on measuring quantum states. The outcomes of measurements affect the flow of the program and determine what code is to be executed next. This has the effect of still using a classical conditional statement, since once qubits are measured they can be considered classical (unless measured in a non-diagonal basis).

Another approach, which is the focus of this article, is the approach taken within the programming languages {QCL} \cite{QCL} and {Silq} \cite{Silq}. The conditional expressions within these languages use boolean expressions to determine what states should act as control qubits and the values those controls activate on. This allows for conditional operators to be described at a higher level than what would be possible using quantum gates.

The problem faced with these conditional statements is that even though gates can be created from specific boolean expressions, \textit{how can an arbitrary boolean function be used to control a quantum operation}?

For instance, consider the oracle in Grover's algorithm, which requires a function $f$ that returns $1$ on a marked element $m$ and $0$ otherwise. The matrix of the oracle for $f$, denoted $G_f$, has a close representation to the identity since $G_f \ket{m}\ket{-} = -\ket{m}\ket{-}$ and $G_f \ket{k}\ket{-} = \ket{k}\ket{-}$ for $k \neq m$. But what happens to the matrix of the oracle when, firstly, there are multiple marked elements and, secondly, when a different unitary operation is used for the oracle? 

The main result of this article is a matrix representation for controlled operations based on an arbitrary function. One can use a common formula to determine the matrix of a controlled gate based on the possible control values and the unitary operations to be performed.

The paper is structured as follows.
Section~\ref{sec:prelim} introduces a background on quantum computing and notation used throughout.
Section~\ref{sec:arb_ctrl_gates} provides definitions and a few properties on gates that are used for this investigation.
Section~\ref{sec:main_res} shows the main result of the paper 
and Section~\ref{sec:applications} discusses potential use cases.
Finally, Section~\ref{sec:conclusion} provides a conclusion.

\section{Preliminaries \label{sec:prelim}}
\subsection{Quantum Computing Background}
This section introduces some concepts of quantum computing required for this paper (notably using Dirac notation). For a full introduction see \cite{NielsenChuang}.

A Hilbert space, denoted $\hil$, is a vector space equipped with an inner product.
Quantum bits (or qubits) are modelled by vectors that span a Hilbert space. A qubit is often represented as
\begin{equation}
    \ket{\phi} = \alpha\ket{0} + \beta\ket{1}
\end{equation}
where $\ket{0} = (1, 0)^T$, $\ket{1} = (0,1)^T$, $\alpha, \beta \in \mathbb{C}$ and $\abs{\alpha}^2 + \abs{\beta}^2 = 1$.
A qubit is said to be in \textit{superposition} if both $\alpha$ and $\beta$ are non-zero, \ie we cannot tell if a qubit is definitely in the state $\ket{0}$ or $\ket{1}$.

The dual of a quantum state is denoted $\bra{\phi} = \ket{\phi}^\dagger$, where $\dagger$ denotes the conjugate transpose of a matrix (the matrix ${A}^\dagger$ has elements ${A_{ij}}^\dagger = \overline{A_{ji}}$).

Hilbert spaces can be expanded by using a tensor product, denoted $\otimes$. This allows us to define systems with multiple qubits.
As an example, for a two qubit system, we have the following state
\begin{equation}
\begin{aligned}
    \bigg(\frac{1}{\sqrt{2}}\ket{0} + \frac{1}{\sqrt{2}}\ket{1}\bigg) \otimes \bigg(\frac{1}{\sqrt{2}}\ket{0} + \frac{1}{\sqrt{2}}\ket{1}\bigg) \\
    = \frac{1}{2} \ket{00} + \frac{1}{2} \ket{01} + \frac{1}{2} \ket{10} + \frac{1}{2} \ket{11}
\end{aligned}
\end{equation}
where $\ket{xy}$ denotes $\ket{x} \otimes \ket{y}$ for $x,y \in \bs{}$.

\begin{figure}[t]
    \centering
     \begin{subfigure}[b]{0.3\textwidth}
         \centering
         \begin{quantikz}
         & \gate{X} & \qw 
         \end{quantikz}
         \caption{NOT/Pauli-X gate}
         \label{sfig:H}
     \end{subfigure}
     \hfill
     \begin{subfigure}[b]{0.3\textwidth}
         \centering
         \begin{quantikz}
         & \ctrl{1} & \qw \\
         & \targ{} & \qw
         \end{quantikz}
         \caption{CNOT gate (CX)}
         \label{sfig:cnot}
     \end{subfigure}
     \hfill
     \begin{subfigure}[b]{0.3\textwidth}
         \centering
         \begin{quantikz}
         & \ctrl{1} & \qw \\
         & \ctrl{1} & \qw \\
         & \targ{} & \qw
         \end{quantikz}
         \caption{Toffoli gate (CCX)}
         \label{sfig:tof}
     \end{subfigure}
    \caption{Quantum circuit diagram for each of the example gates given}
    \label{fig:gates}
\end{figure}
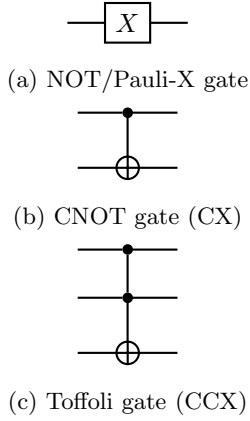

A \emph{unitary} operation $U$ on a quantum system (Hilbert space) is a matrix such that $U^{-1} = U^\dagger$. An example of a single qubit operation is the NOT gate, denoted X, with matrix representation
\begin{equation*}
    X =
    \begin{pmatrix}
    0 & 1 \\
    1 & 0
    \end{pmatrix}.
\end{equation*}

Of importance for this paper are the CNOT gate and the Toffoli gate, denoted CX and CCX respectively. The CNOT gate acts on two qubits, not changing the state of the first qubit, but flipping the state of the second qubit if the first qubit is in the state $\ket{1}$. The matrix representation of a CNOT gate is

\begin{equation*}
CX =
    \begin{pmatrix}
    1 & 0 & 0 & 0 \\
    0 & 1 & 0 & 0 \\
    0 & 0 & 0 & 1 \\
    0 & 0 & 1 & 0
    \end{pmatrix}.
\end{equation*}

Similarly, the Toffoli gate act on three qubits. It does not change the first two qubits, but does flip the final qubit if the first two qubits are in the state $\ket{1}\ket{1}$. The matrix representation of the Toffoli gate is given in Equation~\ref{eq:ccnot} and the circuit representation of these examples is shown in Figure~\ref{fig:gates}.

\small{
\begin{equation}
\label{eq:ccnot}
CCX = \begin{pmatrix}
    1 & 0 & 0 & 0 & 0 & 0 & 0 & 0 \\
    0 & 1 & 0 & 0 & 0 & 0 & 0 & 0 \\
    0 & 0 & 1 & 0 & 0 & 0 & 0 & 0 \\
    0 & 0 & 0 & 1 & 0 & 0 & 0 & 0 \\
    0 & 0 & 0 & 0 & 1 & 0 & 0 & 0 \\
    0 & 0 & 0 & 0 & 0 & 1 & 0 & 0 \\
    0 & 0 & 0 & 0 & 0 & 0 & 0 & 1 \\
    0 & 0 & 0 & 0 & 0 & 0 & 1 & 0
    \end{pmatrix}
\end{equation}
}

\subsection{Notation}
Here notation that is used throughout the paper is introduced:
\begin{itemize}
    \item $\hil$ represents a Hilbert space and $\hil^n$ is a tensor product of $n$ copies of $\hil$.
    \item $[a \twodots b]$: represents the set of integers between $a$ and $b$ inclusively (\ie $\{a, a+1, \twodots, b-1, b\}$).
    \item $[a \twodots b) = [a \twodots b - 1]$
    \item For a positive integer $n$, $N = 2^n$ (similarly for $m$)
    \item For an integer $x \in [0 \twodots N)$, $\ket{x}$ denotes the quantum state $\ket{x_{n-1} \dots x_1 x_0} \in \hil^n$, where $x_{n-1} \dots x_1 x_0$ is the $n$-bit representation of $x$.
    \item The inner product of quantum states $\ket{\phi}, \ket{\psi} \in \hil^n$ is denoted by $    \braket{\phi}{\psi} = [\phi_0^*, \dots, \phi_{N-1}^*] 
    \begin{bmatrix}
        \psi_0 \\
        \vdots \\
        \psi_{N-1}
    \end{bmatrix}$.
    \item $I_n$ represents the identity matrix of size $n \times n$.
    \item $\prod_{i \in \{i_1, \dots, i_n\}} U_i = U_{i_1}\dotproduct U_{i_2} \dots U_{i_{n-1}}\dotproduct U_{i_n}$.
    For the purposes of this paper, it is assumed that $U_k$ is commutative, with respect to dot product, to $U_l$ for any $k, l \in \{i_1, \dots, i_n\}$ (\ie $U_k \dotproduct U_l = U_l \dotproduct U_k$). Generally, this is not the case for arbitrary unitary matrices.
\end{itemize}

\section{Arbitrarily Controlled Quantum Gates \label{sec:arb_ctrl_gates}}
Here we introduce the gates that are of interest and prove a few properties about them.
Let $\hilsys = \hil^n \otimes \hil^m$ (where $n, m$ are positive integers) be a Hilbert space that is spanned by a quantum system.
Throughout, let $U : \hil^m \to \hil^m$ be a unitary operator, $\ket{\phi} \in \hil^m$ and $x \in \bsn$.

\subsection{Binary Controlled Gates (BCGs)}
Firstly, we introduce the concept of a \emph{binary controlled gate}.

\begin{definition}
(Binary Controlled Gate)
Let $y \in \bsn$. Define $CU_y : \hilsys \to \hilsys$ as

\begin{equation}
    CU_y \ket{x} \ket{\phi} = \begin{cases}
    \ket{x} U\ket{\phi} & \text{if $x = y$} \\
    \ket{x} \ket{\phi} & \text{otherwise} \\
    \end{cases}
\end{equation}
\end{definition}

This is a controlled operation that states that if the first $n$ qubits are of the form $y$, then the operation $U$ should be performed on the remaining $m$ qubits.

\begin{lemma}
\label{lem:bcgmat}
Let $CU_y$ be a BCG. Then, we have that
\begin{equation}
\label{eq:cuy_form}
\begin{aligned}
    & \mel{i}{CU_y}{j} = \\
    & \begin{cases}
    \mel{i-y M}{U}{j-y M} & \text{if $i, j \in [yM \twodots (y + 1) M)$} \\
    \delta_{ij} & \text{otherwise}
    \end{cases}
\end{aligned}
\end{equation}
\end{lemma}

\ifx\appendproofs\undefined
\begin{proof}

Firstly, let $\ket{i} = \ket{i'}\ket{s}$ and $\ket{j} = \ket{j'}\ket{r}$, where $i', j' \in \bsn$ and $r, s \in [0 \twodots M)$.
Note that $\bra{i} = \bra{s} \bra{i'}$ and that $i = i'M + s$ (similarly for $j$, $j'$ and $r$ respectively).
By making use of the definition of a BCG, we have that
\begin{equation}
\begin{aligned}
    & \mel{i}{CU_y}{j} = \bra{i}CU_y \ket{j'}\ket{r}\\
    &= \begin{cases}
    \bra{s}\bra{i'} \ket{j'} U \ket{r} & \text{if $j' = y$} \\
    \bra{s}\bra{i'} \ket{j'} \ket{r} & \text{otherwise}
    \end{cases}
\end{aligned}
\end{equation}

The second case is simply $\bra{s}\bra{i'} \ket{j'} \ket{r} = \braket{i}{j} = \delta_{ij}$, so we can focus on the first case.

Let $i' = j' = y$. Then we have that $\braket{i'}{j'} = 1$. Further, through rearrangement $s = i - i'M = i - yM$. When $i' = j' = y$, then the values of $i$ and $j$ are restricted to those in $[yM \twodots (y+1)M)$. Therefore we have that
\begin{equation}
    \label{eq:cuy_pen}
\begin{aligned}
    &\mel{i}{CU_y}{j} = \\
    &\begin{cases}
    \mel{i-y M}{U}{j-y M} & \text{if $i, j \in [yM \twodots (y + 1) M)$} \\
    \bra{s}\bra{i'} \ket{j'} U \ket{r} & \text{if $j' = y, i' \neq j'$} \\
    \delta_{ij} & \text{otherwise}
    \end{cases}
\end{aligned}
\end{equation}

To deal with the case when $i' \neq j'$, note that of $\bra{i'}\ket{j'} = 0$ and $i \notin [yM \twodots (y+1)M)$. Therefore, $\bra{s}\bra{i'} \ket{j'} U \ket{r} = 0 = \delta_{ij}$, as $i \neq j$, and $i$ is not within the interval in the first condition given in Equation~\ref{eq:cuy_pen}. Hence, the result in Equation~\ref{eq:cuy_form} is reached.

\end{proof}
\fi

Using the lemma just shown, we can verify that BCGs are valid quantum operations (\ie BCGs are unitary).
\begin{lemma}
\label{lem:bcgunitary}
Let $CU_y$ be a BCG. Then $CU_y$ is unitary.
\end{lemma}

\ifx\appendproofs\undefined
\begin{proof}

Note that $U$ is unitary and therefore ${U U^\dagger = U^\dagger U = I_M}$.

Firstly, we will show that $(CU_y)^\dagger = C(U^\dagger)_y$, where $C(U^\dagger)_y$ is a BCG using $U^\dagger$ as its controlled operation. By making use of Lemma~\ref{lem:bcgmat}, we have that

\begin{equation*}
\begin{split}
    & \mel{i}{(CU_y)^\dagger}{j} = \mel{j}{\overline{CU_y}}{i}
    \\
    = & \begin{cases}
    \mel{j-y M}{\overline{U}}{i-y M} & \text{if $i, j \in [yM \twodots (y + 1) M)$} \\
    \overline{\delta_{ji}} & \text{otherwise}
    \end{cases}
    \\
    = & \begin{cases}
    \mel{i-y M}{U^\dagger}{j-y M} & \text{if $i, j \in [yM \twodots (y + 1) M)$} \\
    \delta_{ij} & \text{otherwise}
    \end{cases}
    \\
    = & \mel{i}{C (U^\dagger)_y}{j}
\end{split}
\end{equation*}

By the definition of a BCG, we have

\begin{equation}
    C(U^\dagger)_y \ket{x} \ket{\phi} = \begin{cases}
    \ket{x}U^\dagger \ket{\phi} & \text{if $x=y$}\\
    \ket{x}\ket{\phi} & \text{otherwise}
\end{cases}
\end{equation}

Again, by using the definition of a BCG, we have that
\begin{equation*}
\begin{aligned}
    & (C(U^\dagger)_y) CU_y \ket{x}\ket{\phi} 
    \\
    &= \begin{cases}
    C(U^\dagger)_y \ket{x} (U\ket{\phi}) & \text{if x = y} \\
    C(U^\dagger)_y \ket{x}\ket{\phi} & \text{otherwise}
    \end{cases}
    \\
    & = \begin{cases}
    \ket{x} U^\dagger U\ket{\phi} & \text{if x = y} \\
    \ket{x}\ket{\phi} & \text{otherwise}
    \end{cases}
    \\
    & = \ket{x} \ket{\phi}.
\end{aligned}
\end{equation*}

Therefore $(CU_y)^\dagger CU_y = I_{NM}$ and so $CU_y$ is unitary.
\end{proof}
\fi

\begin{exmp}
\label{ex:cx}
Let $n = m = 1$, $U = X$ and $y = 1 \in \{0,1\}$. Then, by following the result from Lemma~\ref{lem:bcgmat}, we have, as expected, that 

\begin{equation*}
CU_y = \begin{pmatrix} 1 & 0 & 0 & 0 \\ 0 & 1 & 0 & 0\\ 0 & 0 & 0 & 1 \\ 0 & 0 & 1 & 0 \end{pmatrix}= CX.    
\end{equation*}
\end{exmp}

\begin{exmp}
Now consider the same set up to Example~\ref{ex:cx} except $n = 2$ and $y = 3 \in [0 \twodots 3]$. By using Lemma~\ref{lem:bcgmat}, the matrix representation of $CU_y$ is the same as that as the Toffoli gate in Equation~\ref{eq:ccnot}.
\end{exmp}

\subsection{Function Controlled Gates (FCG)}
We want to extend the notion of BCGs so that we can perform the same operation on multiple binary values. This is where we introduce the notion of \emph{function controlled gates}.

\begin{definition} (Function Controlled Gate)
Let $f: \bsn \to \{0,1\}$. Define $CU_f: \hilsys \to \hilsys$ as

\begin{equation}
    CU_f \ket{x} \ket{\phi} = \begin{cases}
    \ket{x} U\ket{\phi} & \text{if $f(x) = 1$} \\
    \ket{x} \ket{\phi} & \text{otherwise} \\
    \end{cases}
\end{equation}
\end{definition}

The following lemma shows that we can construct an FCG using BCGs.

\begin{lemma} \label{lem:fcgprod}
Let $\fdomain = \{y \in \bsn : f(y) = 1\}$. Then, we have that
\begin{equation}
    \label{eq:fcgprod}
    CU_f = \prod_{y \in \fdomain} CU_y
\end{equation}
\end{lemma}

\ifx\appendproofs\undefined
\begin{proof}
Set $x$ such that $f(x) = 0$. Therefore, we have that $CU_f \ket{x} \ket{\phi} = \ket{x} \ket{\phi}$.

For all $y \in \fdomain$, we have ${CU_y \ket{x}\ket{\phi} = \ket{x} \ket{\phi}}$.
It is then not hard to see that
\begin{equation}
\label{eq:lcuf=cuy}
\begin{aligned}
(\prod_{y \in \fdomain} CU_y) \ket{x}\ket{\phi} &= \ket{x}\ket{\phi} \\
&= CU_f \ket{x} \ket{\phi}
\end{aligned}
\end{equation}

Now consider the case when $x \in \fdomain$. We have
\begin{align*}
CU_x \ket{x} \ket{\phi} = \ket{x} U\ket{\phi}\\
CU_f \ket{x}\ket{\phi} = \ket{x} U\ket{\phi}
\end{align*}
and that $CU_y \ket{x} \ket{\phi} = \ket{x} \ket{\phi}$ for $y \in \fdomain \setminus \{x\}$.

Therefore
\begin{align*}
(\prod_{y \in \fdomain} CU_y) \ket{x} \ket{\phi} &= \ket{x} U\ket{\phi}\\
&= CU_f \ket{x}\ket{\phi}
\end{align*}
and so we have the result in Equation (\ref{eq:fcgprod}).
\end{proof}
\fi

Since BCGs are unitary operations and an FCG is a product of BCGs, it trivially follows that an FCG is unitary.
\begin{corollary}
Let $CU_f$ be an FCG. Then $CU_f$ is unitary.
\end{corollary}

\begin{exmp}
Consider the oracle gate of $f$, denoted $O_f$, where $O_f \ket{x} \ket{y} = \ket{x} \ket{y \oplus f(x)}$. By adding an ancillary qubit and recalling that $O_f^{-1} = O_f$, an FCG can be simulated by an oracle of $f$ and controlled unitary gate with a control on one qubit (denoted $CU$):

\begin{equation*}
\begin{aligned}
    & \ket{x} \ket{0} \ket{\phi} \\
    & \xrightarrow[]{O_f \otimes I_M} \ket{x} \ket{f(x)} \ket{\phi}\\
    & \xrightarrow[]{I_N \otimes CU} \begin{cases}
    \ket{x} \ket{f(x)} U \ket{\phi} & \text{if $f(x) = 1$} \\
    \ket{x} \ket{f(x)} \ket{\phi} & \text{otherwise}
    \end{cases} \\
    & \xrightarrow[]{O_f \otimes I_M} \begin{cases}
    \ket{x} \ket{0} U \ket{\phi} & \text{if $f(x) = 1$} \\
    \ket{x} \ket{0} \ket{\phi} & \text{otherwise}
    \end{cases}
\end{aligned}    
\end{equation*}
\end{exmp}

\begin{exmp} \label{exmp:oracle}
With FCGs, the phase of a qubit can also be changed based on $f$. This is particularly useful for the phase oracle of different algorithms, such as Grover's algorithm \cite{Grover96}. Consider a function for Grover's algorithm ${f: \bsn \to \{0,1\}}$ that has multiple marked elements, whose set we denote ${M_f \subseteq \bsn}$ (\ie $m \in M_f$ if $f(m) = 1$).

The oracle of $f$ is normally defined as

\begin{equation*}
    G_f \ket{x} = \begin{cases}
    - \ket{x} & \text{if }x \in M_f \\
    \ket{x} & \text{otherwise}
    \end{cases}.
\end{equation*}

However, this normally requires a controlled gate acting on an ancillary qubit to be implemented.\footnote{see page 180 (Chapter 4.3) of \cite{NielsenChuang}}
Here we can use an FCG, in this case taking $f$ as the function and $U = X$:

\begin{align*}
    CU_f \ket{x} \ket{-} & = \begin{cases}
    \ket{x} X \ket{-} & \text{if } x \in f(x) = 1 \\
    \ket{x} \ket{-} & \text{otherwise}
    \end{cases} \\
    & = \begin{cases}
    - \ket{x} \ket{-} & \text{if } x \in M_f \\
    \ket{x} \ket{-} & \text{otherwise}
    \end{cases} \\
    & \equiv G_f \ket{x}
\end{align*}
where $\ket{-} = \frac{1}{\sqrt{2}} (\ket{0} - \ket{1})$.

Even though this example is fairly simple, it can be generalised to handle different phases and using different unitary operations as well.
\end{exmp}

\section{Matrix Representation of Function Controlled Gates \label{sec:main_res}}
In this section, the main result of the article is shown and examples are given to highlight how it can be modified. Firstly, we show the matrix representation of FCGs.

\begin{theorem}\label{thm:main}
Let $CU_f$ be a FCG. Then

\begin{equation}
\begin{aligned}
    \mel{i}{CU_f}{j} =
    \delta_{ij} + f(y) (\mel{i-y M}{U}{j-y M} - \delta_{ij})
\end{aligned}
\end{equation}
if $i, j \in [yM \twodots (y + 1) M)$ and $y\in \bsn$.
Otherwise, $\mel{i}{CU_f}{j} = 0$.
\end{theorem}

\ifx\appendproofs\undefined
\begin{proof}
\input{main}
\end{proof}
\fi

Note that an FCG can be represented as a matrix of $M \times M$ matrices. Given the result of Theorem~\ref{thm:main}, these sub-matrices are either filled with $0$'s (if not on the diagonal) or are of the form $I_M + f(x)(U-I_M)$ where $x$ is the current position on the diagonal.\footnote{The matrix form can be seen in \cite{stackexchange}, where only a single qubit is used for the control and target of the operation.}

Theorem~\ref{thm:main} extends a result in \cite{Samoladas2008} by allowing the unitary operation to act on a quantum register instead of a single qubit. The results of Theorem~\ref{thm:main} are similar to the notion of a \emph{quantum if-then} (QIT) operator defined in \cite{Samoladas2008}.
The \emph{quantum if-then} operator $\text{QIT}(F, U)$ is dependent on a single-unitary operation $U$ and the diagonal matrix, $F$, of the results of a boolean function $f$ (\ie $F = diag(f(0), f(1), \dots, f(N)$). Thus, the QIT operator can be written $\text{QIT}(F,U) = F \otimes U + (I_N - F) \otimes I_M$.
Note that the terms are grouped by the identity on the target register rather than by the boolean function.

\begin{exmp}
The classical OR operation on two bits can be simulated on a quantum computer by using two CNOTs and a Toffoli gate as shown in Figure~\ref{fig:qOR}.

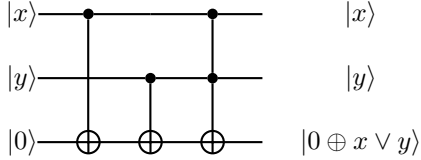
\begin{figure}[ht]
    \centering
    \begin{quantikz}
        \ket{x} & \ctrl{2} & \qw & \ctrl{1} & \qw & \ket{x} \\
        \ket{y} & \qw & \ctrl{1} & \ctrl{1} & \qw & \ket{y} \\
        \ket{0} & \targ{} & \targ{} & \targ{} & \qw & \ket{0 \oplus x \lor y}
    \end{quantikz}
    \caption{An OR operation using quantum gates}
    \label{fig:qOR}
\end{figure}

However, a FCG can be used instead. By setting $U = X$ and using ${OR : [0 \twodots 3] \to \bs{}}$, where ${OR(0) = 0}$ and ${OR(x) = 1}$ otherwise, then

\begin{equation}
    CX_{OR} = CCX \dotproduct (I \otimes CX) \dotproduct (C \otimes I \otimes X)
\end{equation}

\end{exmp}

\begin{exmp} \label{exmp:oracle2}
Continuing from Example~\ref{exmp:oracle}, the oracle can further be represented using an FCG but without an ancillary qubit.
Again consider

\begin{equation*}
    G_f \ket{x} = \begin{cases}
    - \ket{x} & \text{if }x \in M_f \\
    \ket{x} & \text{otherwise}
    \end{cases}.
\end{equation*}

Then we have that

\begin{align*}
    \bra{i} G_f \ket{j} & =
    \begin{cases}
        (-1)^{f(y)} & \text{if } i = j, y \in [0..N) \\ & \text{ such that } i, j \in [y..2y)\\
        0 & \text{otherwise}
    \end{cases} \\
    & = \delta_{ij} (1 - 2f(y)) \\
    & = \delta_{ij} + f(y) (-\braket{i}{j} - \delta_{ij})\\
    & = \delta_{ij} + f(y) (\bra{i} (-1) \ket{j} - \delta_{ij})\\
\end{align*}

This form is very close to that of a FCG, with the exception that instead of using a unitary operator, we simply need the phase factor of the oracle.

\end{exmp}

\section{{Applications} \label{sec:applications}}
The main application of our work is in two areas: simulation of quantum circuits and verification.


For simulation, one advantage is that the user can represent an arbitrary controlled operation using a single FCG rather than a complex series of smaller controlled operations. This also has the additional benefit of providing a single operation to execute, rather than multiple operations that need to be executed.


Similarly, this matrix can be used in verification to check that a specified quantum circuit meets its intended behaviour. By using this format, we now have an easy way to reason about the function for an algorithm. Considering an implementation of Grover's algorithm, we can specify the behaviour of $f$ as

\begin{equation*}
    \exists m \text{ such that } \forall x, (f(x) = 1 \equiv x = m)
\end{equation*}
and encode in appropriate syntax for different tools.

Alongside the recent developments of quantum software, some progress is being made in verifying programs for quantum software. An overview of the tools currently available can be found in \cite{Lewis21,Chareton21}.
Some of these verification tools use reductions of the Theorem~\ref{thm:main}.

Isabelle Marries Dirac \cite{Bordg2021, Isabelle_Marries_Dirac-AFP} proves various mathematical properties of quantum circuits using the Isabelle theorem prover \cite{isabelle-hol}. The definitions of \textit{deutsch-transform} and \textit{jozsa-transform}, which are matrices of Deutsch and Deutsch-Jozsa oracles respectively, follow the form given in Example~\ref{exmp:oracle2}.

The form used in those definitions could be used in QHLProver \cite{QHLProver-AFP}, another verification tool based in Isabelle and based on Quantum Hoare Logic \cite{YingFHL}, in the definition of \textit{mat-O} (which gives the matrix form of the Grover oracle).


\begin{exmp}
To give an example of this technique being used, consider the Silq code given in Figure~\ref{fig:silq}. A conditional statement is used with both an if statement, notably with the boolean expression $f(x)$, and an else statement.

\begin{figure}[th]
    \centering
    \begin{minted}[linenos]{d}
    def H[n:!N](x:uint[n]){
        for i in [0..n){ x[i] := H(x[i]); }
        return x;
    }
    
    def example(f : uint[3] -> B){
        x := H[3](0:uint[3]);
        y := 0:uint[2];
        if f(x) {
            y := H[2](y);
        }
        else{
            y[0] := X(y[0]);
            y[1] := X(y[1]);
        }
        ...
    }
    \end{minted}
    \caption{A simple Silq program that uses a conditional statement with a quantum variable in the boolean expression. The condition in the \textit{if} statement is shorthand for the boolean expression $f(x) == 1$.}
    \label{fig:silq}
\end{figure}

Note that in Theorem~\ref{thm:main}, the identity gate is used to consider that there is no operation performed if the quantum state does not meet the condition of $f$. However, by replacing the identity operator with the operation in the else statement, we can achieve a matrix representation of a full conditional (if-then-else) statement.

Thus, by considering the matrix form of Theorem~\ref{thm:main} and changing the identity (represented by $\delta_{ij}$) to the operator within the else-statement, we can consider the gate representing the operation in Figure~\ref{fig:silq} to be
\begin{equation*}
    \begin{pmatrix}
    A(0) & 0 & \dots & 0 \\
    0 & A(1) & \dots & 0 \\
    \vdots & \vdots & \ddots & \vdots \\
    0 & 0 & \dots & A(7) \\
    \end{pmatrix}
\end{equation*}
where $A(x) = X^{\otimes 2} + f(x)(H^{\otimes 2} - X^{\otimes 2})$.
\end{exmp}

\section{Conclusion \label{sec:conclusion}}
In this article, two types of controlled gates were introduced: binary controlled gates and function controlled gates. We showed properties about these gates and provided examples to give motivation for their use. Specifically, the relation between function controlled gates and quantum conditional statements (such as those used in Silq~\cite{Silq}) has been demonstrated.

These gates can provide a useful way to easily create gates in simulators or for purposes of verifying quantum programs, requiring the user only to specify a function and unitary operation to perform. Further, these types of gates give a higher level method as to how controlled gates can be constructed, moving away from using circuit descriptions for conditional statements where it is not immediately obvious what a circuit is doing.


\bibliographystyle{acm}
\bibliography{refs.bib}

\if\appendproofs1
\appendix
\section{Proofs}
\subsection{Proof of Lemma~\ref{lem:bcgmat}}

Firstly, let $\ket{i} = \ket{i'}\ket{s}$ and $\ket{j} = \ket{j'}\ket{r}$, where $i', j' \in \bsn$ and $r, s \in [0 \twodots M)$.
Note that $\bra{i} = \bra{s} \bra{i'}$ and that $i = i'M + s$ (similarly for $j$, $j'$ and $r$ respectively).
By making use of the definition of a BCG, we have that
\begin{equation}
\begin{aligned}
    & \mel{i}{CU_y}{j} = \bra{i}CU_y \ket{j'}\ket{r}\\
    &= \begin{cases}
    \bra{s}\bra{i'} \ket{j'} U \ket{r} & \text{if $j' = y$} \\
    \bra{s}\bra{i'} \ket{j'} \ket{r} & \text{otherwise}
    \end{cases}
\end{aligned}
\end{equation}

The second case is simply $\bra{s}\bra{i'} \ket{j'} \ket{r} = \braket{i}{j} = \delta_{ij}$, so we can focus on the first case.

Let $i' = j' = y$. Then we have that $\braket{i'}{j'} = 1$. Further, through rearrangement $s = i - i'M = i - yM$. When $i' = j' = y$, then the values of $i$ and $j$ are restricted to those in $[yM \twodots (y+1)M)$. Therefore we have that
\begin{equation}
    \label{eq:cuy_pen}
\begin{aligned}
    &\mel{i}{CU_y}{j} = \\
    &\begin{cases}
    \mel{i-y M}{U}{j-y M} & \text{if $i, j \in [yM \twodots (y + 1) M)$} \\
    \bra{s}\bra{i'} \ket{j'} U \ket{r} & \text{if $j' = y, i' \neq j'$} \\
    \delta_{ij} & \text{otherwise}
    \end{cases}
\end{aligned}
\end{equation}

To deal with the case when $i' \neq j'$, note that of $\bra{i'}\ket{j'} = 0$ and $i \notin [yM \twodots (y+1)M)$. Therefore, $\bra{s}\bra{i'} \ket{j'} U \ket{r} = 0 = \delta_{ij}$, as $i \neq j$, and $i$ is not within the interval in the first condition given in Equation~\ref{eq:cuy_pen}. Hence, the result in Equation~\ref{eq:cuy_form} is reached.

\subsection{Proof of Lemma~\ref{lem:bcgunitary}}

Note that $U$ is unitary and therefore ${U U^\dagger = U^\dagger U = I_M}$.

Firstly, we will show that $(CU_y)^\dagger = C(U^\dagger)_y$, where $C(U^\dagger)_y$ is a BCG using $U^\dagger$ as its controlled operation. By making use of Lemma~\ref{lem:bcgmat}, we have that

\begin{equation*}
\begin{split}
    & \mel{i}{(CU_y)^\dagger}{j} = \mel{j}{\overline{CU_y}}{i}
    \\
    = & \begin{cases}
    \mel{j-y M}{\overline{U}}{i-y M} & \text{if $i, j \in [yM \twodots (y + 1) M)$} \\
    \overline{\delta_{ji}} & \text{otherwise}
    \end{cases}
    \\
    = & \begin{cases}
    \mel{i-y M}{U^\dagger}{j-y M} & \text{if $i, j \in [yM \twodots (y + 1) M)$} \\
    \delta_{ij} & \text{otherwise}
    \end{cases}
    \\
    = & \mel{i}{C (U^\dagger)_y}{j}
\end{split}
\end{equation*}

By the definition of a BCG, we have

\begin{equation}
    C(U^\dagger)_y \ket{x} \ket{\phi} = \begin{cases}
    \ket{x}U^\dagger \ket{\phi} & \text{if $x=y$}\\
    \ket{x}\ket{\phi} & \text{otherwise}
\end{cases}
\end{equation}

Again, by using the definition of a BCG, we have that
\begin{equation*}
\begin{aligned}
    & (C(U^\dagger)_y) CU_y \ket{x}\ket{\phi} 
    \\
    &= \begin{cases}
    C(U^\dagger)_y \ket{x} (U\ket{\phi}) & \text{if x = y} \\
    C(U^\dagger)_y \ket{x}\ket{\phi} & \text{otherwise}
    \end{cases}
    \\
    & = \begin{cases}
    \ket{x} U^\dagger U\ket{\phi} & \text{if x = y} \\
    \ket{x}\ket{\phi} & \text{otherwise}
    \end{cases}
    \\
    & = \ket{x} \ket{\phi}.
\end{aligned}
\end{equation*}

Therefore $(CU_y)^\dagger CU_y = I_{NM}$ and so $CU_y$ is unitary.

\subsection{Proof of Lemma~\ref{lem:fcgprod}}
Set $x$ such that $f(x) = 0$. Therefore, we have that $CU_f \ket{x} \ket{\phi} = \ket{x} \ket{\phi}$.

For all $y \in \fdomain$, we have ${CU_y \ket{x}\ket{\phi} = \ket{x} \ket{\phi}}$.
It is then not hard to see that
\begin{equation}
\label{eq:lcuf=cuy}
\begin{aligned}
(\prod_{y \in \fdomain} CU_y) \ket{x}\ket{\phi} &= \ket{x}\ket{\phi} \\
&= CU_f \ket{x} \ket{\phi}
\end{aligned}
\end{equation}

Now consider the case when $x \in \fdomain$. We have
\begin{align*}
CU_x \ket{x} \ket{\phi} = \ket{x} U\ket{\phi}\\
CU_f \ket{x}\ket{\phi} = \ket{x} U\ket{\phi}
\end{align*}
and that $CU_y \ket{x} \ket{\phi} = \ket{x} \ket{\phi}$ for $y \in \fdomain \setminus \{x\}$.

Therefore
\begin{align*}
(\prod_{y \in \fdomain} CU_y) \ket{x} \ket{\phi} &= \ket{x} U\ket{\phi}\\
&= CU_f \ket{x}\ket{\phi}
\end{align*}
and so we have the result in Equation (\ref{eq:fcgprod}).

\subsection{Proof of Theorem~\ref{thm:main}}
Using Lemma \ref{lem:fcgprod}, we have that
\begin{equation}
    \label{eq:fcgprodbcg}
    \mel{i}{CU_f}{j} = \mel{i}{\bigg( \prod_{y \in \fdomain} CU_y \bigg) }{j}
\end{equation}.

Given that the values of $y$ are distinct, these reflect affecting different square matrices along the diagonal. Therefore, we have that

\begin{equation}
\begin{aligned}
    &\mel{i}{\bigg( \prod_{y \in \fdomain} CU_y \bigg) }{j} = \\
    &\begin{cases}
    \mel{i-yM}{U}{j-yM} & \parbox[t]{.25\textwidth}{if $i, j \in [yM \twodots (y + 1) M)$ for $y \in \fdomain$} \\
    \delta_{ij} & \text{otherwise}
    \end{cases}
\end{aligned}
\end{equation}

Consider when $y \in \fdomain$, so $f(y) = 1$. Then for $i, j \in [yM \twodots (y + 1) M)$ we have that

\begin{equation}
    \label{eq:f1_form}
    \begin{aligned}
        & \mel{i-yM}{U}{j-yM} = \delta_{ij} - \delta_{ij} + \mel{i-yM}{U}{j-yM} \\
        &= \delta_{ij} + f(y) (\mel{i-yM}{U}{j-yM} - \delta_{ij}).
    \end{aligned}
\end{equation}

A similar observation can be made when $y \notin \fdomain$ (\ie when $f(y) = 0$) and $i, j \in [yM \twodots (y + 1) M)$:
\begin{equation}
    \label{eq:f0_form}
    \begin{aligned}
        \delta_{ij} &= \delta_{ij} + 0 (\mel{i-yM}{U}{j-yM} - \delta_{ij}) \\
        &= \delta_{ij} + f(y) (\mel{i-yM}{U}{j-yM} - \delta_{ij}).
    \end{aligned}
\end{equation}

Note that for all indexes where $i=j$ have the form as given in Equations~\ref{eq:f1_form} and~\ref{eq:f0_form} and so $i$ and $j$ must lie in some boundary $[yM \twodots (y + 1) M)$. Therefore, when at least one of $i$ or $j$ is not in the interval, then we have that $\delta_{ij} = 0$.
Thus
\begin{equation*}
\begin{aligned}
    \mel{i}{\bigg( \prod_{y \in \fdomain} CU_y \bigg) }{j}
    & = \delta_{ij} + \\
    & f(y) (\mel{i-yM}{U}{j-yM} - \delta_{ij})
\end{aligned}
\end{equation*}
if $i, j \in [yM \twodots (y + 1) M)$ for $y \in \bsn$ and $0$ otherwise.
\fi

\end{document}